\documentclass[%
 reprint,
superscriptaddress,
preprintnumbers,
nofootinbib,
 amsmath,amssymb,
 aps,
]{revtex4-1}

\usepackage[colorlinks]{hyperref}
\usepackage{graphicx}
\usepackage{dcolumn}
\usepackage{bm}
\usepackage{color}
\usepackage{placeins}
\usepackage{slashed}
\usepackage{multirow}
\usepackage{subfigure}
\usepackage{tabularx}
\usepackage{mathtools}
\usepackage{floatrow}
\usepackage{blindtext}
\usepackage{soul}

\definecolor{coral}{RGB}{255,127,80}
\definecolor{indigo}{RGB}{75,0,130}
\definecolor{red}{rgb}{0.9, 0,0}
\definecolor{cerulean}{rgb}{0., 0.62,0.9}
\definecolor{navy}{rgb}{0.05, 0.05,0.8}

\newfloatcommand{capbtabbox}{table}[][\FBwidth]
\renewcommand{\eqref}[1]{Eq.~\ref{#1}}

\begin{document}

\title{
Searching for axion-like particles with data scouting at ATLAS and CMS \\
}

\author{Simon Knapen}
\email{smknapen@lbl.gov}
\affiliation{CERN, Theoretical Physics Department, Geneva, Switzerland}
\affiliation{Berkeley Center for Theoretical Physics, Department of Physics, University of California, Berkeley, CA 94720, USA}
\affiliation{Theoretical Physics Group, Lawrence Berkeley National Laboratory, Berkeley, CA 94720, USA}
\author{Soubhik Kumar}
\email{soubhik@berkeley.edu}
\affiliation{Berkeley Center for Theoretical Physics, Department of Physics, University of California, Berkeley, CA 94720, USA}
\affiliation{Theoretical Physics Group, Lawrence Berkeley National Laboratory, Berkeley, CA 94720, USA}
\author{Diego Redigolo}
\email{d.redigolo@gmail.com}
\affiliation{CERN, Theoretical Physics Department, Geneva, Switzerland}
\affiliation{INFN, Sezione di Firenze Via G. Sansone 1, 50019 Sesto Fiorentino, Italy}

\preprint{CERN-TH-2021-216}

\date{\today}

\begin{abstract}
{
We investigate the physics case for a dedicated trigger-level analysis for very low mass diphoton resonances at ATLAS and CMS and introduce a new photon isolation criterion. This greatly increases the acceptance for light particles at the expense of a somewhat larger Standard Model background.  We show how such an analysis would likely yield new experimental coverage for axion-like particles for masses below 15 GeV, bridging the gap with the region covered by flavor factories.
}
\end{abstract}

\maketitle

\tableofcontents

\section{Introduction}
Diphoton resonances have long been recognized as a fertile ground for potential new discoveries, most notably by the discovery of the Standard Model (SM) Higgs boson in the $\gamma\gamma$ channel \cite{CMS:2012qbp,ATLAS:2012yve}. The latter was of course anticipated before the construction of the LHC, and as a result both ATLAS and CMS have excellent photon identification and reconstruction capabilities. On the theory front, it is well known that any new scalar or pseudoscalar particle with a coupling  charged fermions will typically generate a coupling to two photons at the one loop level. As was the case for the SM Higgs, this naively subleading channel can be a primary discovery mode if the decay widths to the SM fermions are suppressed or background limited. These considerations have resulted in a number of powerful searches, so far with sensitivity in the mass range between 65 GeV and 3 TeV \cite{ATLAS:2014jdv,CMS:2018dqv,CMS:2018cyk,ATLAS:2021uiz}.  
Extending the reach to lower invariant masses is feasible in the near future with present data and proposals have already been put forward both for ATLAS and CMS~\cite{Mariotti:2017vtv,CidVidal:2018eel} as well as LHCb \cite{CidVidal:2018blh}.\footnote{Both ATLAS and CMS have carried out a search for lighter diphoton resonances in ultraperipheral heavy ion collisions~\cite{CMS:2018erd,ATLAS:2020hii}. These searches are however only sensitive if the branching ratio of the new state to the $\gamma\gamma$ channel is $\mathcal{O}(1)$ \cite{Knapen:2016moh}.} In view of future LHC runs, it is important to investigate how much further new trigger techniques could extend the exploration of the diphoton spectrum at low invariant masses.   

The diphoton invariant mass can be written as
\begin{align}\label{eq:massestimate}
m_{\gamma\gamma}&=\sqrt{2p_{T_{1}}^{\gamma}p_{T_{2}}^{\gamma}(\cosh\Delta\eta_{\gamma\gamma}-\cos\Delta\phi_{\gamma\gamma})}\nonumber\\
&\simeq\sqrt{p_{T_{1}}^{\gamma}p_{T_{2}}^{\gamma}}\Delta R_{\gamma\gamma}\ ,
\end{align}
where $p_{T_{1}}^{\gamma}$ ($p_{T_{2}}^{\gamma}$) is the transverse momentum of the leading (subleading) photon, and $\Delta R_{\gamma\gamma}=\sqrt{\Delta\eta_{\gamma\gamma}^2+\Delta\phi_{\gamma\gamma}^2}$ is their angular separation in the limit where both pseudorapidity separation $\Delta\eta_{\gamma\gamma}\ll1$ and azimuthal separation $\Delta\phi_{\gamma\gamma}\ll1$. This simple formula is instrumental to understand why searches for low invariant mass resonances face severe challenges related to the trigger system. The primary problem is the huge rate of low $p_T$ photons produced in parton showers and in meson decays, $\pi^0$ decays in particular. In order to keep the rate manageable, minimal $p_T$ thresholds and photon isolation criteria are needed. Given \eqref{eq:massestimate}, these result in a lower bound on the attainable diphoton invariant mass in a given search. 

For example, in~\cite{CMS:2018cyk} CMS relied on a 
trigger which required a $p_{T_{1}}^{\gamma}>30$ GeV ($p_{T_{2}}^{\gamma}>18$ GeV) threshold on the leading (subleading) photon, tight isolation 
and an invariant mass requirement of $m_{\gamma\gamma}>55$ GeV. The latter limited the accessible mass range in the analysis to $m_{\gamma\gamma}>70$ GeV, once the turn-on of the trigger was taken into account. 
To address this issue, ATLAS has implemented a special trigger stream geared towards low mass diphoton resonances~\cite{ATLAS:2019dpa}. At the hardware/level 1 trigger (L1 trigger), it requires two photons with a symmetric \mbox{$p_{T_{1,2}}^{\gamma}>15$ GeV} requirement and loose isolation. This L1 trigger seeds a software/high-level trigger (HLT) with  \mbox{$p_{T_{1,2}}^{\gamma}>22$ GeV}. This trigger combination collected already 138.5 $\text{fb}^{-1}$ data during the past LHC run, and it currently represents the best way of exploring the low invariant mass region with ATLAS. CMS has a similar, but asymmetric trigger with $p_{T_{1}}^{\gamma}>25$ GeV ($p_{T_{2}}^{\gamma}>12$ GeV) L1 thresholds for the leading (subleading) photon, with and without a loose isolation criterion~\cite{CMS:2020cmk}. 

Here we explore an alternative, forward-looking strategy to probe low mass resonances, which we benchmark against the existing ATLAS trigger \cite{ATLAS:2019dpa}. The minimal photon $p_T$ and the diphoton angular separation are interconnected by simple kinematics if we look at diphoton resonances, as in \eqref{eq:massestimate}. For this reason, once the trigger thresholds on the photons are accounted for, at low masses the resonances will be sufficiently boosted such that the two photons can spoil each others isolation criteria. Our proposal tries to address this problem in two ways.

First, we assess whether a strategy of directly using photons reconstructed at the trigger level (called Turbo Stream \cite{Aaij:2019uij}, Data Scouting \cite{CMS:2016ltu} or Trigger-Level Analysis \cite{ATLAS:2018qto}) could yield a further improvement relative to the existing ATLAS trigger \cite{ATLAS:2019dpa}. In such a strategy, only a small fraction of the full event record is written to tape, which allows for a much higher output rate. The ATLAS, CMS and LHCb collaborations have successfully deployed this technique for low-mass resonance searches, in particular in the context of dijet \cite{CMS:2016ltu,ATLAS:2018qto} and dimuon \cite{PhysRevLett.124.131802,CMS-PAS-EXO-20-014,LHCb:2019vmc} resonances.  For the diphoton case, applying this technique could theoretically allow for lower $p_T$ thresholds and possibly less stringent isolation requirements, as compared to the normal trigger stream. 

In addition we investigate whether the tension between the $p_T$ thresholds and isolation cut can be further alleviated by defining a dedicated isolation variable for a diphoton pair: when summing the $p_T$ of the activity within the isolation cone, we will explicitly exclude the hardest, other photon candidate. We also attempt to assess how the background from jets faking photons behaves when subjected to this different isolation requirement. We however do not attempt to modify the shower shape variables and our study can thus be seen as a step towards a broader class of ``photon-jet'' objects at trigger level~\cite{Ellis:2012zp,Allanach:2017qbs,Sheff:2020jyw,Wang:2021uyb,Ren:2021prq}, where the isolation requirements and identification criteria on the shower shape are tailored to the features of the targeted signal.  

The experimental implementation of an analysis like what we propose here is both a very time- and resource-consuming endeavour, and we therefore seek to first estimate its possible gain with a phenomenological study. Our results will hopefully aid the experimental collaborations in determining whether a diphoton analysis directly at the trigger level is feasible and desirable, given the resources at their disposal. 

To quantify the possible gain in sensitivity, we use a well motivated, KSVZ style \cite{Kim:1979if,Shifman:1979if} axion-like particle (ALP) as an example benchmark model, defined by
\begin{equation}
\mathcal{L}_{a}\supset -\frac{1}{2}m_a^2 a^2- \frac{\alpha_s}{8\pi}\frac{a}{f_a} G\tilde{G}+ \frac{E}{N}\frac{\alpha_{\text{em}}}{8\pi}\frac{a}{f_a} F\tilde{F}\,.\label{eq:ALPL}
\end{equation}
Here $N$ and $E$ are the anomaly coefficients for the gluon and photon couplings respectively and the $a\to \gamma\gamma$ branching ratio is specified by the ratio of $E/N$.  
In KSVZ models which are compatible with grand unification one typically finds \mbox{$E/N\sim\mathcal{O}(1)$}, resulting in \mbox{Br($a\to\gamma\gamma)\sim 10^{-3}$} and \mbox{Br($a\to jj)\sim 1$}. Despite this seemingly small branching ratio, the $\gamma\gamma$ channel is still a superior signature due to the huge dijet background. The gluon coupling on the other hand can yield very large cross sections for $m_a$ below 100 GeV, partially offsetting the small branching ratio to photons.

The remainder of our paper is organized as follows: in Sec.~\ref{sec:simulation} we describe our simulation framework and analysis strategy, followed by a brief summary of the theory motivation for low mass diphoton resonances in Sec.~\ref{sec:theory}. We conclude in Sec.~\ref{sec:conclusions}. In Appendix~\ref{app:appendix} we provide further details on our simulation framework.

%
%
%
\section{Analysis\label{sec:simulation}}
In this section we describe our simulation framework and analysis strategy, with a special emphasis on how our isolation criteria are defined. We take the ATLAS calorimeter as a primary example, but most aspects of our study should carry over to CMS. 

\subsection{Photon identification and isolation\label{sec:iso}}
In ATLAS Run-2 data taking the photon candidate has to satisfy reconstruction, identification and isolation requirements which are described in \cite{ATL-PHYS-PUB-2011-007,ATLAS:2019qmc,ATLAS:2009zsq}. Here we briefly review the main steps, with a focus on how these requirements can limit the signal efficiency for low mass diphoton resonances. First of all, the photon selection at L1 is based on a reduced ECAL granularity. At present, a trigger ECAL tower is taken to be a rectangle of $\Delta\eta\times\Delta\phi=0.075\times 0.098$~\cite{ATLAS:2016ecu}.\footnote{The actual granularity of a single ECAL cell is $\Delta\eta\times\Delta\phi=0.025\times 0.0245$ in the middle layer and goes down to $\Delta\eta=0.0031$ if we consider the strip towers in first layer of the ATLAS ECAL. This more refined information is however only available at the HLT.}  

Offline, a photon candidate is classified as a (loosely) reconstructed photon if a number of criteria are satisfied: first, the energy deposited in a $0.25\times 0.25$ window of the hadronic calorimeter (HCAL) behind the ECAL energy cluster has to be very small compared to the energy deposited in the ECAL. (The ratio between HCAL and ECAL energy deposits is called hadronic leakage in the experimental literature.) Second, ATLAS imposes requirements on the shape of the electromagnetic shower in the middle ECAL layer, such as an upper bound on the lateral width of the shower. In addition they define the variables 
\begin{equation}
R_\eta \equiv \frac{E_{3\times 7}}{E_{7\times 7}}\quad \mathrm{and} \quad R_\phi  \equiv \frac{E_{3\times 3}}{E_{3\times 7}}.
\end{equation}
Here, e.g.,~$E_{3\times 7}$ is the energy deposited in a $3\times 7$ block of ECAL cells, centered around the barycenter of the photon candidate, where the cell are organized in an $\eta\times \phi$ grid. Both $R_\eta$ and $R_\phi$ are bounded from above, to ensure the shower of the photon candidate is sufficiently contained. The above variables are used for loose reconstruction and their primary role is to reject photons from collimated $\pi^0$ decays~\cite{ATL-PHYS-PUB-2011-006,ATL-PHYS-PUB-2011-007,ATLAS:2018fzd}. 


Loose (tight) photon isolation furthermore imposes an upper bound on the amount of extra energy deposit within a cone of $\Delta R<0.2$ ($\Delta R<0.4$) around the photon candidate. In particular, the ATLAS loose photon isolation variable is defined as
\begin{equation}
r_{\gamma}\equiv\frac{1}{p_{T_\gamma}}\Bigg(\!\!\!\sum_{\substack{\Delta R<0.2\\ p_{T_i}>0.5 \rm{GeV}}}\!\!\!\!\! p_{T_i}\Bigg)\ ,\label{eq:isoATLAS}
\end{equation}
and similarly for tight isolation. A photon is considered isolated if $r_{\gamma}<0.05$.\footnote{In other studies, such as Ref.~\cite{Catani:2018krb}, isolation is defined as an upper bound on the total transverse energy inside the isolation cone, $E_{T}^{\text{MAX}}=r_{\gamma}p_{T_\gamma}$.} The sum runs over all tracks and calorimeter deposits with $p_T>0.5$ GeV and a distance $\Delta R<0.2$ from the photon candidate, excluding the candidate itself. This isolation criterion is meant to veto photon candidates that originated from inside a jet, but it can also reject genuine, light diphoton resonances when combined with the trigger requirements on the photon $p_T$.

To address this problem, we apply a trigger-level photon isolation criterion where the photons are not counted towards each other's isolation variable, as will be explained in~\eqref{eq:modiso}. We do not modify the other reconstruction and identification criteria. As such, we still require that the minimal separation between a diphoton pair is such that two L1 trigger ECAL towers do not overlap, which roughly corresponds to $\Delta R_{\gamma\gamma}\gtrsim 0.1$. Of course, a more detailed study by the experimental collaborations is required to properly assess the behavior of the reconstruction and identification algorithms in combination with the modified isolation defined here. For instance, the use of the ATLAS $R_\eta$ and $R_\phi$ variables as defined now may not be compatible with our modified isolation.

For our simulations, we rely on the \texttt{Delphes 3}  framework \cite{deFavereau:2013fsa}, which propagates the particles of the event through a simplified detector volume and assigns a fraction of the particle's energy to a particular collection of calorimeter cells. The candidate photons after this process are identified with those ECAL cells for which there is no significant energy deposit in the corresponding tower in the hadronic calorimeter and for which there is no track pointing towards the ECAL cell. 

To determine which photon candidates qualify as reconstructed photons, \texttt{Delphes 3} uses the same isolation criterion as in \eqref{eq:isoATLAS}. Candidates which fail isolation are included among the inputs for the jet clustering algorithm \cite{Cacciari:2011ma}. 
%
Since we wish to preserve photon candidates which have another hard photon candidate within their isolation cone, we modify the standard isolation procedure by explicitly subtracting the $p_T$ of the hardest photon candidate in the isolation cone from the sum. Concretely, we modified the \texttt{Delphes 3} isolation module to define a new isolation variable $r_{\gamma\gamma}$, which is more suitable for photons which are expected to come in pairs
\begin{equation}\label{eq:modiso}
r_{\gamma\gamma}\equiv\frac{1}{p_{T_\gamma}}\Bigg(\!\!\!\sum_{\substack{\Delta R<0.2\\ p_{T_i}>0.5 \rm{GeV}}}\!\!\!\!\! p_{T_i}-p_{T_{\gamma,1}}\Bigg),
\end{equation}
where $p_{T_\gamma}$ is the transverse momentum of the primary photon candidate, and $p_{T_{\gamma,1}}$ is the transverse momentum of the leading photon candidate in the isolation cone, excluding the primary photon candidate itself. The sum is defined as in \eqref{eq:isoATLAS}. A photon candidate is then classified as an isolated photon if it satisfies $r_{\gamma\gamma}<0.05$. A similar modification of the isolation variable was already used in an offline analysis~\cite{ATLAS:2015rsn}.  

The isolation criterion described in~\eqref{eq:isoATLAS} relies on tracking information. However, tracking is a computing-intensive step, and it may therefore not be available at high enough rates in an analysis that includes photons reconstructed directly at the trigger level, due to the limited resources available in the trigger farm. In the context of diphoton resonances, tracking primarily improves the efficacy of pile-up subtraction methods and isolation cuts, leading to a lower number of fake photons.\footnote{Tracking is also  important to extract data driven estimates for the photon fake rate~\cite{ATLAS:2011gau}. If tracking is not available for the main trigger stream, a separate prescaled stream may be needed to obtain the required control samples.} 
We therefore expect larger backgrounds for an analysis which cannot rely on tracking. To estimate how much a search would degrade in the absence of tracking when computing the isolation, we also repeat the procedure outlined above while setting the tracking efficiency for all charged particles to zero. 

\subsection{Analysis cuts\label{sec:analysis}}

As the implementation of our signal and the background simulations are informed by our fiducial cuts, we first discuss those here. We must make a number of assumptions, in particular in relation to which $p_T$ cuts may be viable for a trigger-level analysis. For guidance, we look at the ATLAS L1 thresholds for two electromagnetic objects \cite{ATLAS:2019dpa}, which where set to $p_T>10$ GeV and $p_T>15$ GeV in the 2015 and 2016 data sets respectively (\texttt{L1\_2EM10VH} and \texttt{L1\_2EM15VH}). As of the 2017 data set, a loose isolation requirement was added  (\texttt{L1\_2EM15VHI}). We will here optimistically consider $p_T>15$ GeV but assume no isolation requirement at L1, effectively mimicking the cuts of the \texttt{L1\_2EM15VH} trigger item. 
We assume that the modified isolation criteria described in the previous subsection would be implemented online after the L1 trigger.
The rate for the \texttt{L1\_2EM15VH} selection was measured to be roughly \mbox{5 kHz} \cite{ATLAS:2019dpa}, which is similar to what was recorded for an existing trigger-level analysis \cite{CMS-PAS-EXO-20-014}. 

We estimate the expected sensitivity for the following three cases:
\begin{itemize}
\item[(I)] Trigger-level with tracking information, modified isolation (\eqref{eq:modiso}), and requiring $p_{T\gamma}>15$~GeV for each photon, denoted by `Tracking'. 
\item[(II)] Same as above but without tracking, denoted by `No tracking'.
\item[(III)] As a stand-in for an offline analysis with the existing ATLAS trigger, we require $p_{T\gamma}>25$~GeV for each photon \cite{ATLAS:2019dpa}.\footnote{We choose $p_{T\gamma}>25$~GeV, as compared to the $p_{T\gamma}>22$~GeV in the ATLAS HLT trigger strategy~\cite{ATLAS:2019dpa}, to approximately take into account the turn-on of the trigger.}
 For this case we assume standard isolation (\eqref{eq:isoATLAS}) with tracking information. 
Going forward, we will refer to this as the `Offline' case.

\end{itemize}

Each event is required to contain two isolated photons with pseudorapidity cut $|\eta_\gamma| < 2.5$, each above the $p_T$ threshold as mentioned above.


\begin{figure}
\centering
\includegraphics[width=\textwidth]{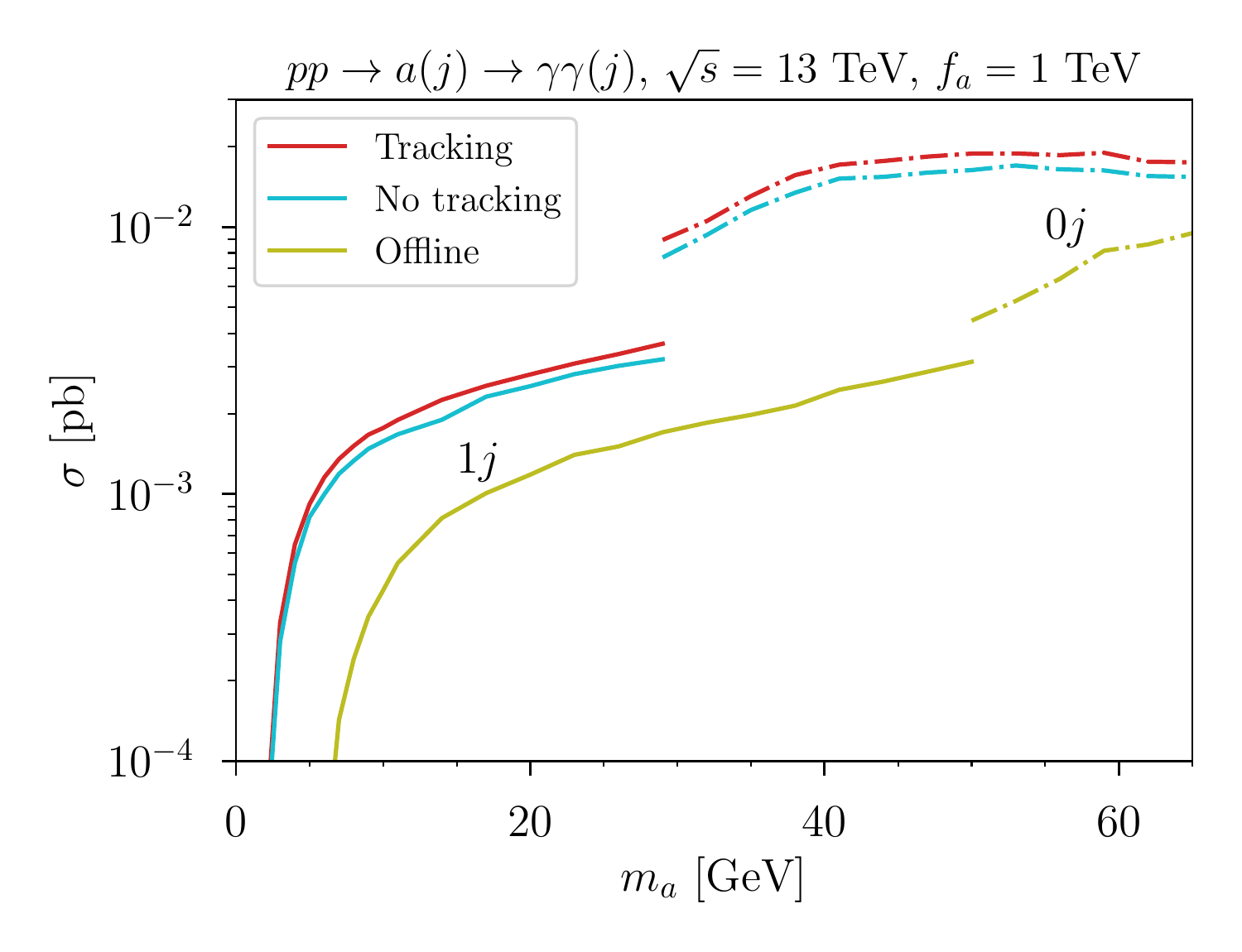}
\caption{Fiducial cross sections for the $a+$jet (solid) and $a$ (dot-dashed) inclusive cross sections, for the selections described in the text. We switch between the two samples when $m_a$ is twice the photon $p_T$ threshold. The discontinuity can be resolved as a smooth step transition by matching the samples at next-to-leading order, which we did not attempt here. For \mbox{$m_a\lesssim 10$ GeV}, the photons are too collimated to pass the isolation criterion for the ``offline'' selection.}  
\label{fig:sig_eff}
\end{figure}

\subsection{Signal}
The ALP benchmark model was implemented in Feynrules \cite{Alloul:2013bka}, to facilitate the generation of Monte Carlo sample with \texttt{Madgraph 5} \cite{Alwall:2011uj} at leading order. The showering and hadronization was done with \texttt{Pythia 8} \cite{Sjostrand:2006za, Sjostrand:2007gs}. We assumed a KSVZ ALP with vector-like matter in a set of $5$-$\bar 5$ multiplets, compatible with $SU(5)$ unification. This choice corresponds to $E/N=8/3$ in relation to~\eqref{eq:ALPL} and is only relevant for the $a\to\gamma\gamma$ branching ratio. The $a\to\gamma\gamma$ branching ratio is sensitive to NNLO corrections to gluon width~\cite{Chetyrkin:1998mw} and has therefore a mild dependence on $m_a$. For $5\,\text{GeV}< m_a < 100\, \text{GeV}$, it is well approximated by the phenomenological formula
\begin{align}\label{eq:branchingratio}
\text{Br}&(a\to\gamma\gamma)\approx 7.8\times10^{-4}\times \left(\frac{E/N}{8/3}\right)^2 \nonumber\\
&\times\left[1+0.67 \log\left(\frac{m_a}{10\,\text{GeV}}\right)
+0.12 \log\left(\frac{m_a}{10\,\text{GeV}}\right)^2\right].
\end{align}

The detector response was simulated with \texttt{Delphes 3}, with the standard ATLAS card, which we  modified to more accurately represent the angular resolution of the ATLAS ECAL (see Appendix~\ref{app:resolution}). We assumed an average of 40 pile-up collisions per event and employed basic pile-up mitigation strategies, as built into \texttt{Delphes 3}. Concretely, all tracks with a longitudinal separation of more than $0.1$ mm from the primary vertex are removed and a pile-up density correction is applied to the photon isolation variable \cite{deFavereau:2013fsa}.

 At low $m_a$, the most relevant process is an ALP produced in gluon fusion together with an associated jet. The latter is needed to ensure that the photons from the ALP decay can pass the trigger $p_T$ thresholds. The $a$ + jet production process was modeled at the level of the hard matrix element, using  \texttt{Madgraph 5} with a minimal cut on the jet $p_T$: \mbox{$p_{T,j}>30$~GeV} as in~\cite{Hook:2019qoh}. The fiducial rate does not increase if a fully inclusive (matched) sample is used instead, as long as $m_a$ is less than twice the photon $p_T$ threshold. We also verified that matching up to two jets does not significantly change the relevant differential distributions.  Following \cite{Gershtein:2020mwi}, we then compute the $a + 1 j$ cross section at the NLO with Madgraph@NLO~\cite{Alwall:2014hca,Artoisenet:2013puc}.  The fiducial cross sections subject to the cuts specified in Sec.~\ref{sec:analysis} are shown in Fig.~\ref{fig:sig_eff} (solid lines).

For $m_a$ larger than twice the photon $p_T$ threshold no recoiling jet is needed to satisfy the trigger requirements. An inclusive signal sample without demanding an additional jet is therefore more suitable, as indicated by the dot-dashed lines in Fig.~\ref{fig:sig_eff}. The sample was rescaled to match the NNLO cross section in \cite{Mariotti:2017vtv}, which was computed with the \texttt{ggHiggs v3.5} package \cite{Ball:2013bra,Bonvini:2014jma,Bonvini:2016frm}.  A smooth interpolation between both regimes can be achieved by matching the zero jet and one jet sample at NLO, which we did not attempt here.
 
The main take-away from Fig.~\ref{fig:sig_eff} is that the ``offline'' selection starts to loose acceptance for $m_a\lesssim$ 15 GeV, whereas both the ``tracking'' and ``no tracking'' selections maintain reasonable acceptance as low as $m_a\approx 5$ GeV. We also see from Fig.~\ref{fig:sig_eff} that the signal efficiency does not degrade much in the absence of tracking, as expected. 

\subsection{Background \label{sec:background}}
\begin{figure*}
    \centering
    \includegraphics[width=0.33\textwidth]{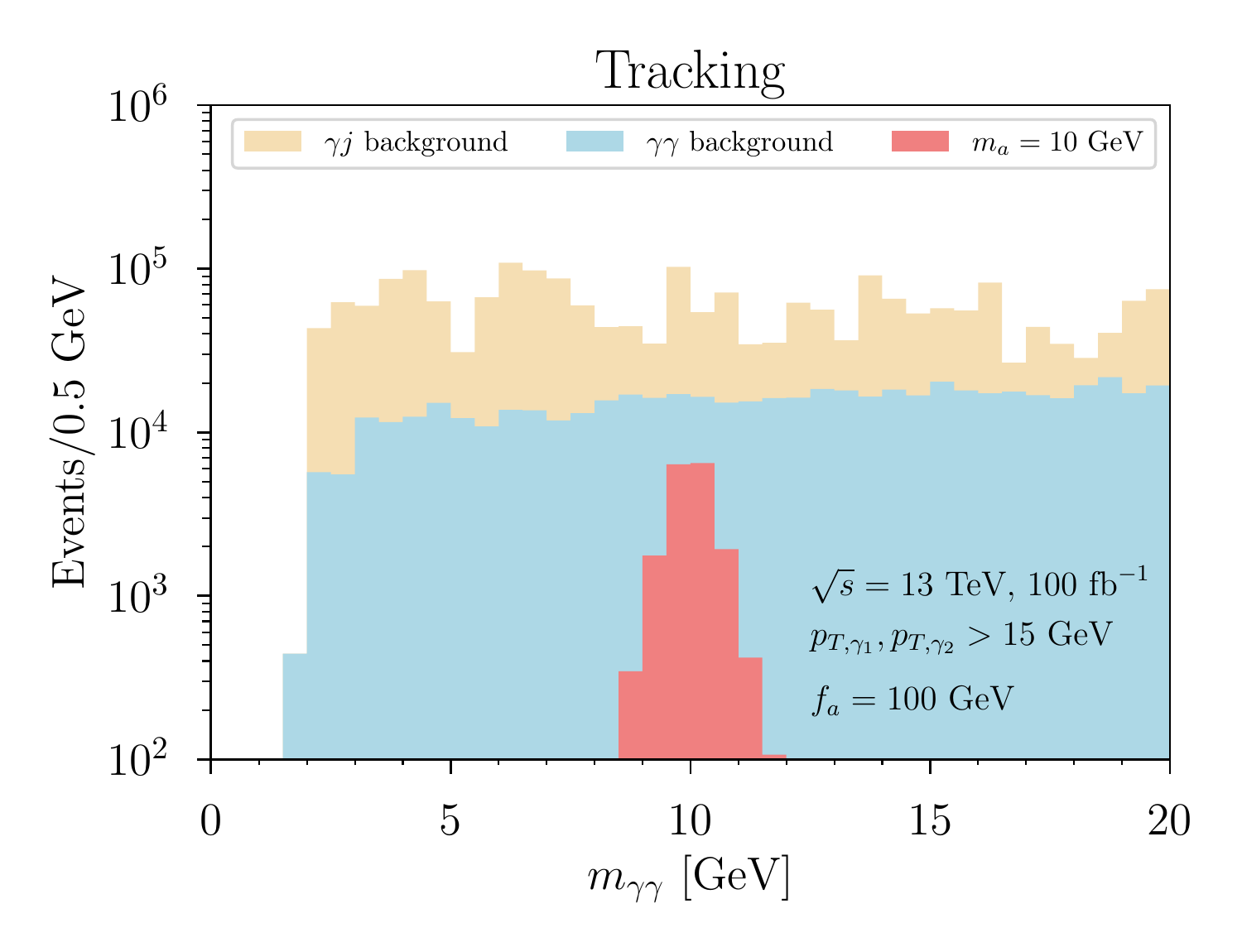}\hfill
    \includegraphics[width=0.33\textwidth]{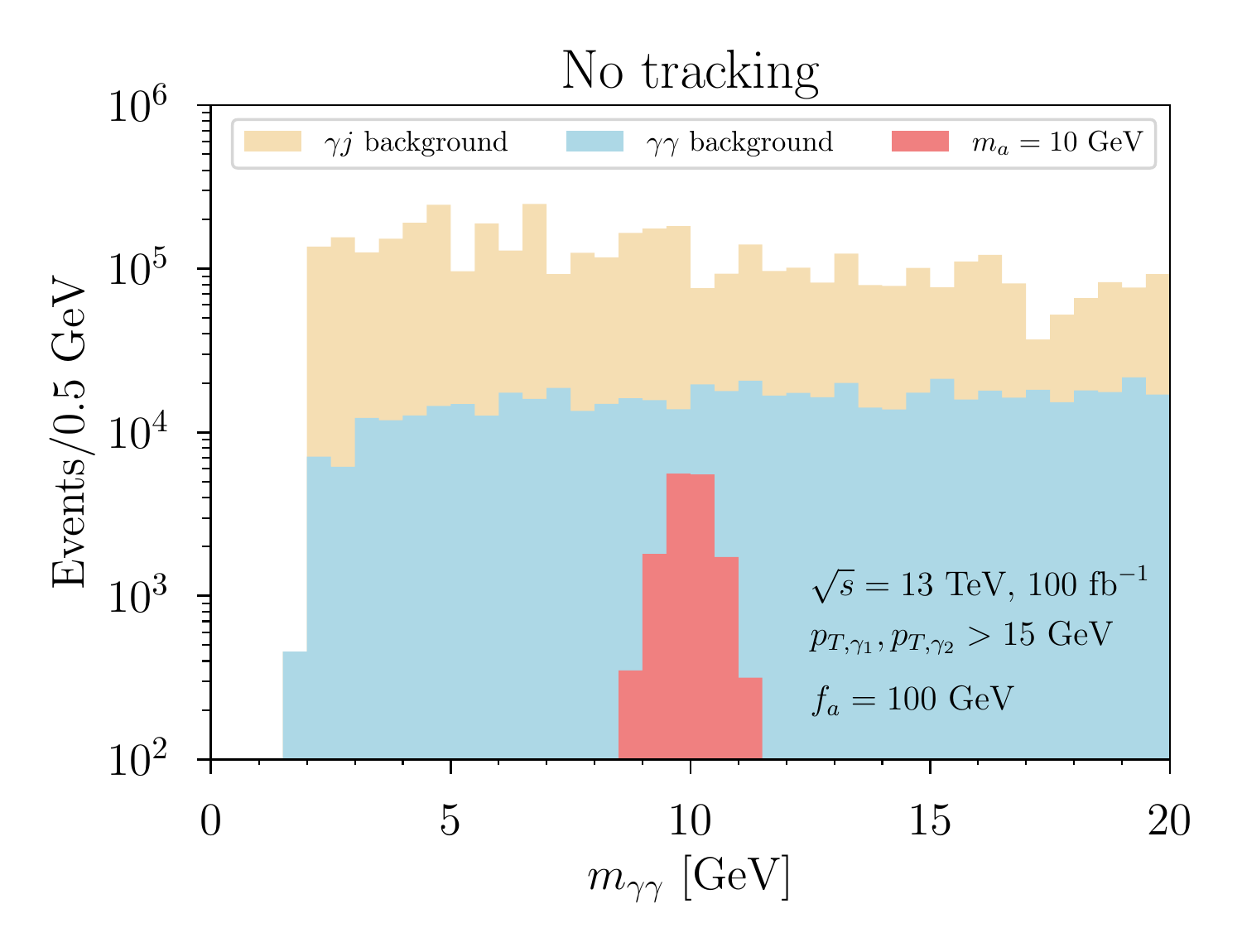}\hfill
    \includegraphics[width=0.33\textwidth]{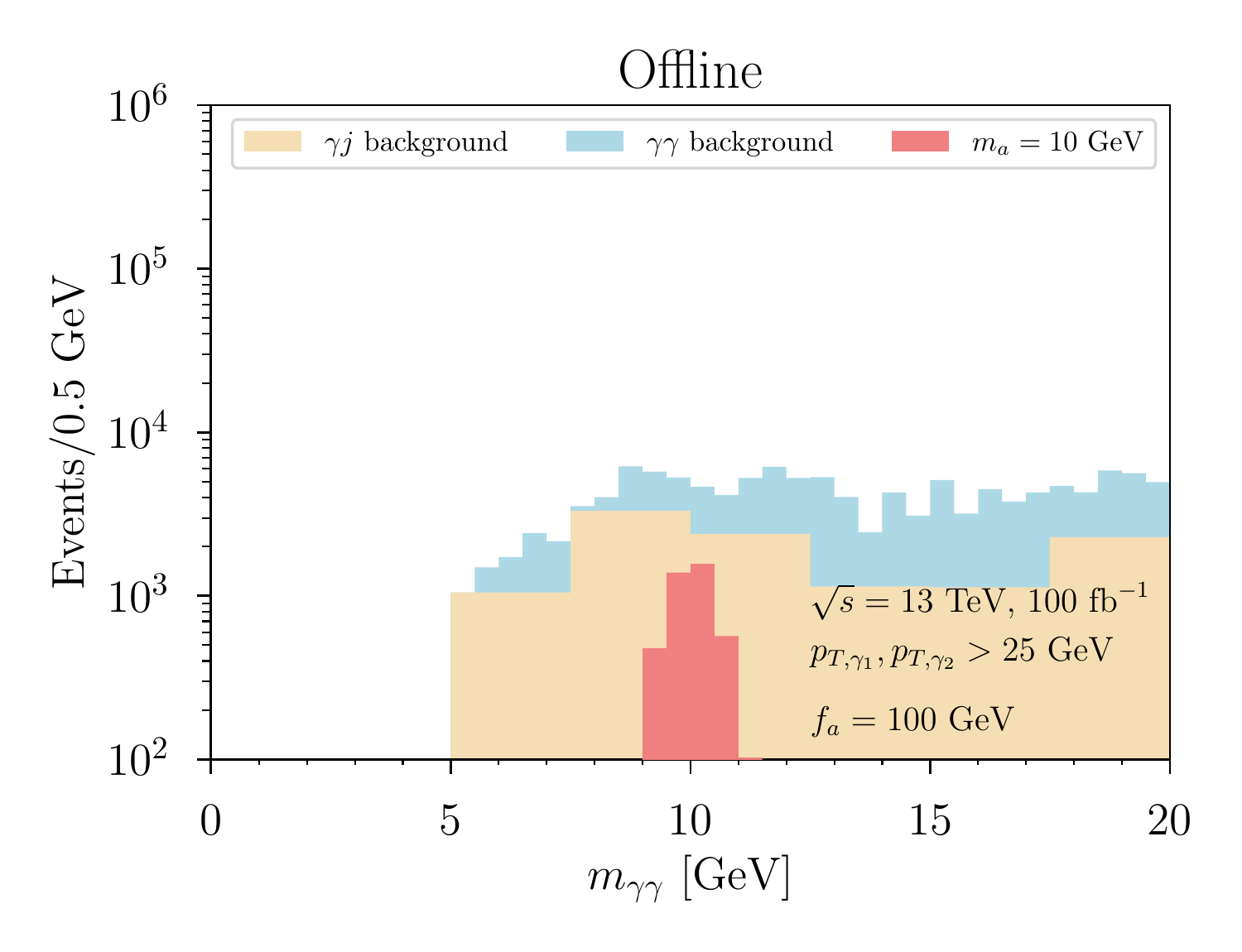}
    \caption{Stacked invariant mass distribution for the signal and the background for the three scenarios described in Sec.~\ref{sec:analysis}. A representative signal resonance is shown for $m_a=10$~GeV with total number of events corresponding to $f_a=100$~GeV. For the same plot with a larger mass range, see Fig.~\ref{fig:inv_mass_dist_app} in the appendix. The `$\gamma j$ background' in the panel labeled `Offline' uses larger bin widths due to
    more limited simulation statistics. See Sec.~\ref{sec:background} for details.
       \label{fig:inv_mass_dist}
       }
\end{figure*}

One of main backgrounds are events with pairs of relatively collimated photons, which recoil against a jet with moderate $p_T$. 
In the SM, photons arise either from the hard scattering matrix element, emission of a fermion line in the parton shower or in hadron decays, notably $\pi^0$ decays. Isolation criteria are designed to single out the former contribution, which give us the most insight in the short distance physics. The shower and hadronization contributions are captured in the perturbative and non-perturbative components of the photon fragmentation functions, e.g.~see \cite{Bourhis:1997yu}, and are ideally counted towards their corresponding jet. This is achieved to a very high degree with the use of suitable isolation criteria. However, even very small fake probabilities (probability that an isolated photon arises through showering or hadronization) can be important due to the comparatively high dijet and photon plus jet cross sections. This means that backgrounds from jets faking photons cannot a priori be neglected.

We opt to simulate all components with a combination of \texttt{Madgraph 5} and \texttt{Pythia 8}, as described below. We validate our simulation against the NNLO calculation by Catani et~al.~\cite{Catani:2018krb}, as well as experimental data \cite{ATLAS:2011gau} (see also \cite{ATLAS:2021mbt} for a recent update at $\sqrt{s}=13\text{ TeV}$). This is described in Appendix~\ref{app:background}. 

Photons originating from the hard collision were modeled with an inclusive $\gamma\gamma$ sample, matched up to two jets. 
The events were then passed through \texttt{Delphes 3} as described in the previous sections. The collinear photon-quark singularity was regulated with an angular cut of $\Delta R_{q\gamma}>0.05$ at the generator level. The overall cross section was rescaled to the NLO inclusive $\gamma\gamma $ cross section as computed with Madgraph@NLO. The resulting invariant mass distributions are shown in Fig.~\ref{fig:inv_mass_dist}, labeled as ``$\gamma\gamma$ background'', for our three different benchmark strategies, as described in Sec.~\ref{sec:analysis}. 

A second background component arises from photon plus jet production, where the jet fakes an additional photon. The most dominant contribution is the $qg\to q\gamma$ process, whenever an additional hard photon is emitted during the fragmentation of the outgoing quark. Fragmentation photons may be identified as isolated if they are radiated at a wide angle from their corresponding quark line. Such fake photons are relatively rare, however it is essential to estimate their rate, because the $\gamma$ plus jet cross section is much larger than the $\gamma\gamma$ cross section. 
This is partially due to the additional $\alpha_{\rm em}$ suppression of the latter, but also due to luminosity functions: at leading order, the $\gamma\gamma$ process requires a $q\bar q$ initial state, whose luminosity function is much smaller than the $qg$ luminosity. 

\begin{figure}
\includegraphics[width=\textwidth]{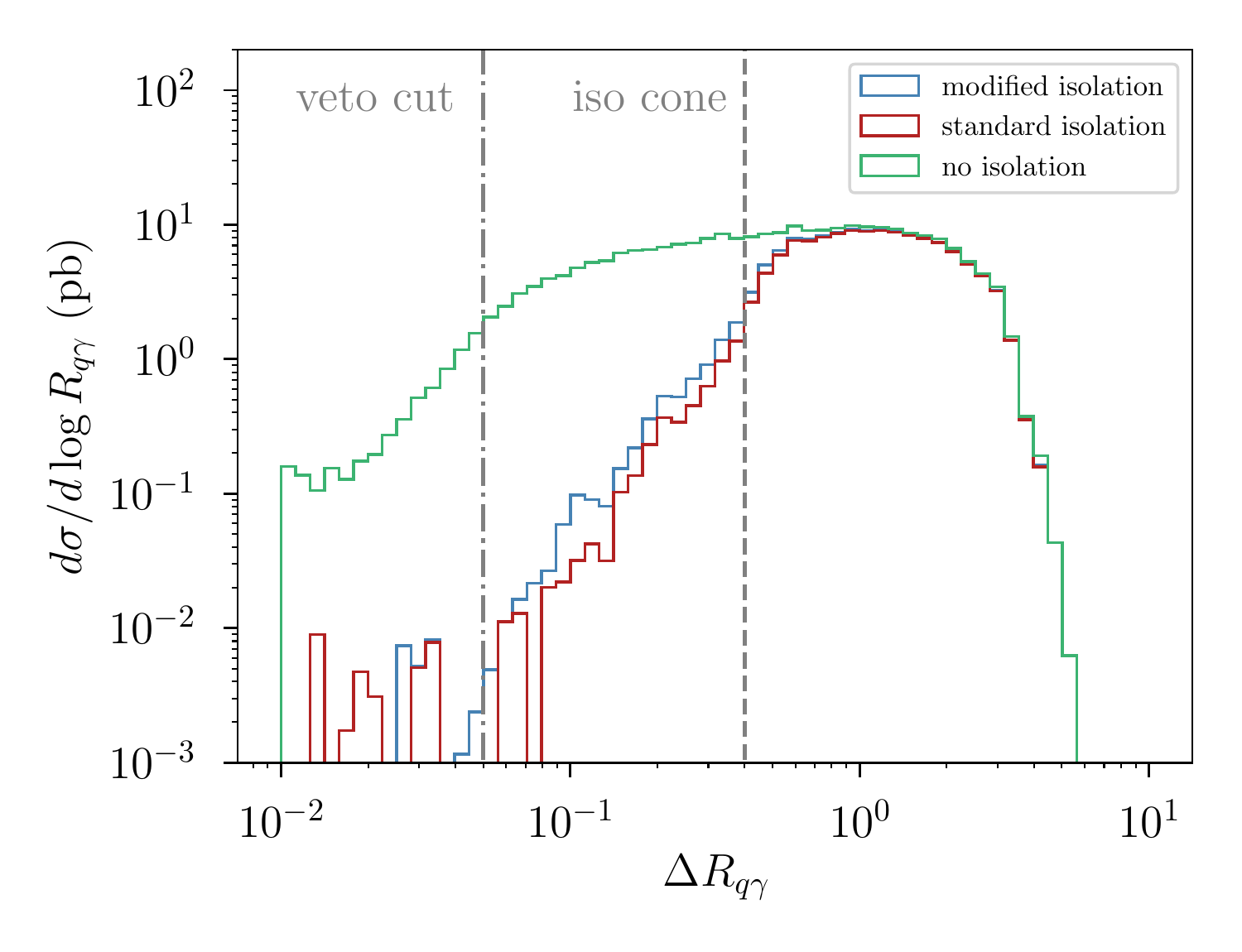}
\caption{ Angular separation of hard ($p_T>15$ GeV) fragmentation photons from their sister quark in a leading order $gq\to q \gamma$ sample, simulated with \texttt{Pythia 8} and rescaled to match the NLO, inclusive photon plus jet cross section. Only photons from the perturbative parton shower are included. The veto cut and isolation cone are indicated for reference. (See text for details.)
 \label{fig:vetocone}}
\end{figure}

To characterize this contribution it is helpful to look at the kinematics of the $q\to q\gamma$ splitting function: In the colinear limit the angle between the photon and the outgoing quark is given by
\begin{equation}\label{eq:splitting}
\theta_{\gamma q} \approx \frac{Q}{E_q} \sqrt{\frac{1}{z(1-z)}}\ ,
\end{equation}
where $Q$ and $E_q$ are the virtuality and energy of the incoming quark respectively, while $z=E_\gamma/E_q$ is the energy fraction carried away by the photon. The \texttt{Pythia 8} parton shower is ordered according to virtuality, which through \eqref{eq:splitting} implies that wide angle radiation occurs primarily in the initial stages of the parton shower. However, emissions with \mbox{$\Delta R_{q\gamma}>0.05$}  populate the same region of phase space as the hard process simulation described in the previous paragraph. We therefore veto them to avoid double counting. The effect of this veto is shown in Fig.~\ref{fig:vetocone}, where we show the $\Delta R_{q\gamma}$ distribution for fragmentation photons, subjected normal isolation, modified isolation and no isolation. Photons from the non-perturbative parts of the fragmentation process, e.g.,~hadron decays, are accounted for separately and not included in this plot. As expected, all three distributions align for $\Delta R_{q\gamma}$ larger than the isolation cone, while the two isolated distributions drop rapidly for $\Delta R_{q\gamma}<0.4$. Since all photons with $\Delta R_{q\gamma}>0.05$ are veto-ed, we find that this contribution to the isolated photon distribution can be neglected.

Fake photons can also be  produced in the non-perturbative part of the quark fragmentation function, primarily from $\pi^0$ decays. To estimate this part, we simulate a  $pp \to \gamma j (j)$ sample with \texttt{Madgraph 5}, matched up to two jets. The cross section was rescaled to match the inclusive, NLO $\gamma j$ cross section as computed with Madgraph@NLO. The events were subsequently showered and hadronized with \texttt{Pythia 8}. In this step all electromagnetic radiation within the jet was veto-ed, as we separately accounted for this component in the previous paragraph. The resulting distribution is shown in Fig.~\ref{fig:inv_mass_dist}, labeled as ``$\gamma j$ background''. It is comparable or larger than the true photon background for our modified isolation criterion, and smaller than the true photon background for standard isolation. The fake photon background can be further reduced by imposing tighter isolation cuts; we comment on this in Appendix~\ref{app:background}.

In our discussion above we have always assumed that at least one photon originated from the hard matrix element. A priori it is possible that both the photons are supplied by the fragmentation of quarks in multi-jet processes. This was found to be negligible by Catani~et~al.~\cite{Catani:2018krb}, although in a slightly different kinematical regime. We verified that this conclusion extends to our kinematical cuts, as long as the fragmentation probability for both photons are assumed to be uncorrelated. Whether this assumption holds for the low $m_{\gamma\gamma}$ regime of interest must be verified with data.

Finally, in a realistic experimental environment fake photons also arise from electrons for which the track was not reconstructed or from hadrons which stop in the first part of the ECAL. Both are difficult to model with a theorist-level simulation and they are neglected in this study. 

%
%

\subsection{Results\label{sec:results}}
As expected, Fig.~\ref{fig:inv_mass_dist} makes it clear that both signal and background go up substantially for a trigger-level analysis, as compared to the ``offline'' scenario. A trigger-level approach is most clearly useful for $m_a\lesssim 10$ GeV, as here the signal efficiency of the ``offline'' selection dies off rapidly, as seen from Fig.~\ref{fig:sig_eff}. For $m_a\gtrsim 10 $ GeV the relative power of both approaches will depend on a number of important experimental subtleties, such as the precise thresholds and turn-on of the various triggers as well as various sources of systematic uncertainties. 

To produce an estimate for the reach of both strategies, a number of additional assumptions are therefore needed. In particular, we will assume that the dominant uncertainty on the background is statistical rather than systematic. This is a reasonable assumption for a bump-hunt over a smooth background, however, any features in the invariant mass distribution of the diphoton background below the trigger threshold might introduce further complications. We also do not attempt a full fledged likelihood analysis and instead simply compare the number of signal and background events in bins with width set to the expected invariant mass resolution (See Appendix~\ref{app:resolution}). This can be justified because of the large values of expected signal and background events. For example, as can be seen in Fig.~\ref{fig:inv_mass_dist}, the number of background and signal events (for $m_a<100$~GeV and $f_a<1$~TeV) in a single invariant mass bin containing the signal is $\sim 10^4$ and $\sim 10^2$ respectively. One can first construct a test statistic~\cite{Cowan:2010js}
\begin{align}
t = -2\ln\left(\frac{L(n|b)}{L(n|s+b)}\right)  \end{align}
to obtain the significance. Here $L(n|m)$ denotes the probability of observing $n$ events originating from a Poisson distribution with mean $m$,
and $s$ and $b$ denote the expected number of signal and background events respectively. By assuming $n=s+b\gg1$ and expanding in $s\ll b$, $t$ can be written as
\begin{align}
t = 2\left(n\ln\left(\frac{s+b}{b}\right)-s\right) \approx \frac{s^2}{b}.    
\end{align}
In the limit of large number of samples, $t$ is $\chi^2$-distributed with one degree of freedom under the null hypothesis. Therefore, for each value of $m_a$, we consider a single invariant mass bin which contains $90\%$ of the signal and determine $f_a$ such that $\sqrt{t}=2$, to obtain a $2\sigma$ confidence limit.



\begin{figure}
    \centering
    \includegraphics[width=\textwidth]{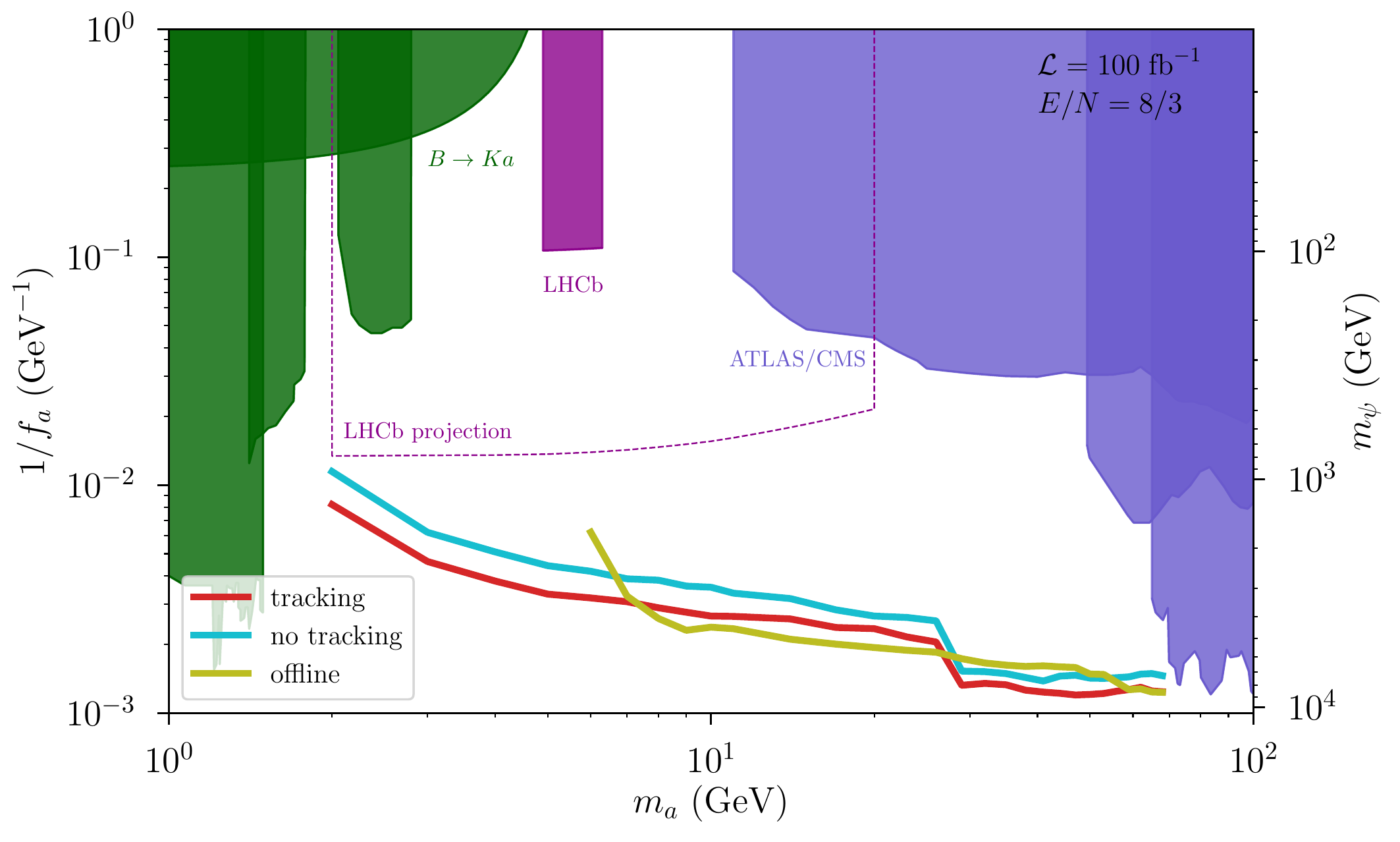}
  \caption{Estimated 2$\sigma$ exclusion reach for 100 $\text{fb}^{-1}$, along with existing constraints from ATLAS \cite{ATLAS:2014jdv,Mariotti:2017vtv}, CMS \cite{CMS:2017dcz,CMS-PAS-HIG-14-037,CMS-PAS-HIG-17-013,CMS:2017dcz,Mariotti:2017vtv}, LHCb \cite{CidVidal:2018blh} and
exotic $B$ to $a$ decays \cite{Chakraborty:2021wda}. The smooth green contour represents the bound on the inclusive $B\to s a$, while the remaining green contours represent several exclusive channels. These bound are somewhat model dependent, see \cite{Chakraborty:2021wda} for details.
We also show an aggressive projection for LHCb, assuming \mbox{300 $\text{fb}^{-1}$} \cite{CidVidal:2018blh}. The right-hand $y$-axis indicates the mass of the colored fermions ($\psi$) for the example UV completion described in Sec.~\ref{sec:validity}.
    \label{fig:projections}}
\end{figure}

With this simplified recipe we find similar sensitivity for all approaches for $m_a\gtrsim 10$ GeV but see that a \mbox{trigger-level} analysis can potentially uniquely probe the $m_a\lesssim10\text{ GeV}$ region, as shown in Fig.~\ref{fig:projections}. While the overall normalization of our estimate is sensitive to all the subtleties listed above, we emphasize that this feature should be robust, as it is driven by the behavior of the signal efficiency in Fig.~\ref{fig:sig_eff}. All strategies would however be a significant improvement over the existing bounds. 

Finally, we comment on the possibility to observe the Standard Model $\eta_b\to \gamma\gamma$ decay.\footnote{We thank Michelangelo Mangano for pointing this out to us.} The $\eta_b$ is the lightest state in the bottomonium system with mass $m_{\eta_b}\approx 9.4$ GeV. For a slightly boosted $\eta_b$ with $p_T\approx 20 $ GeV, its production cross section at the LHC is expected to be between 1 and 10 nb \cite{Likhoded:2017kfw}. The $\eta_b$ branching ratio to photons has not been measured yet but is thought to be roughly $5\times 10^{-5}$\cite{Fabiano:2003yk}. This rate might be high enough to be observable in a trigger-level analysis, depending on the efficiency of our isolation requirements. We leave this interesting question for a future study.

\section{Theory motivation\label{sec:theory}}
We first comment on the generic features of the UV completions of the effective theory in \eqref{eq:ALPL}. Then we briefly review some of the top-down motivation for low mass diphoton resonances.

\subsection{Example UV completion\label{sec:validity}}

For completeness, we should investigate whether there exist healthy UV completions of the effective action in~\eqref{eq:ALPL}, given the values of $f_a$ that are probed in Fig.~\ref{fig:projections} and the complementary searches at the LHC probing new physics at this scale.

As a minimal example we take the model by Kim, Shifman, Vainshtein and Zakharov (KSVZ)~\cite{Kim:1979if,Shifman:1979if}, where the ALP is the phase of a complex scalar singlet $\Phi$.  The model also includes a set of heavy, vector-like colored and electrically charged fermions that are charged under the PQ symmetry. This specifies the action
\begin{equation}\label{eq:ksvz}
\mathcal{L}_{\text{KSVZ}}\supset g_*\Phi \tilde{\psi}\psi,\quad  \quad \Phi=\frac{v_a+\varphi}{\sqrt{2}}e^{ia/v_a}.
\end{equation}
In our conventions, the vacuum expectation value of $\Phi$ ($v_a$) is related to the ALP decay constant by $v_a=2 Nf_a $. The anomaly coefficients $N$ and $E$ are specified by the multiplicity and the representation of the fermions. Taking the $\psi$ to fill out a $5$-$\bar5$ representation of $SU(5)$, we have  $N=N_\psi/2$ and $E=4/3 N_\psi$ with $N_\psi$ the number of $\psi$ flavors. Finally, the $g_*$ coupling determines the mass of the fermions $m_\psi=N_\psi g_*f_a/\sqrt{2}$. Taking for example $N_\psi=5$ and $g_*= 3$ ensures that the $\psi$ states can be well heavier than the reach of the many LHC searches for new colored particles. (See right hand $y$-axis in Fig.~\ref{fig:projections}). This example indicates that the discovery of a low mass diphoton resonance would likely be the hallmark of an elaborate composite sector near the TeV scale.

 Alternatively, it is not difficult to increase the branching ratio to photons in \eqref{eq:branchingratio} by choosing exotic representations which yield a larger $E/N$ ratio. One can also consider low values of $f_a$ but lift the masses of the colored particles in the multiplet by means of an additional mass term in \eqref{eq:ksvz}. This does not reduce the overall signal rate, since the reduced production cross section is exactly compensated by an increase in the branching ratio to photons, as long as the decay width to photons remains small compared to the decay width to gluons. This remains true as long as the mass of the colored particles satisfies $M\lesssim (\alpha_s/\alpha_{\rm em})\, g_\ast f_a$. For the parameters in Fig.~\ref{fig:projections}, this easily pushes the colored states outside the energy reach of the LHC. 

\subsection{More complete models}
The ALP benchmark model defined in \eqref{eq:ALPL} can be the hallmark of more complete models addressing the hierarchy, the strong CP or the dark matter problems. These three motivations were reviewed comprehensively in \cite{CidVidal:2018eel,Gershtein:2020mwi}, and we therefore only summarize the main points.

The main virtue of very low scale supersymmetry breaking is that a viable gravitino abundance can be achieved without the need for a non-standard cosmological history \cite{Osato:2016ixc}. In such scenarios, one also expects its associated $U(1)_R$ symmetry to be broken at a low scale, leading to a parametrically light pseudo-goldstone boson, the $R$-axion \cite{Nelson:1993nf,Intriligator:2007py,Bellazzini:2017neg}. The $R$-axion couples to the gauginos, which induce the operators in \eqref{eq:ALPL} in the low energy theory.  

Axions famously provide a solution to the strong CP problem by promoting the QCD $\theta$-angle to a dynamical field \cite{Peccei:1977hh,Peccei:1977ur}. This mechanism however comes with a \emph{PQ quality problem}, which states that any sources of PQ breaking other than the QCD dynamics must be very small. The suppression of potentially dangerous Planck-suppressed operators can achieved through extending the UV structure of the model~\cite{Izawa:2002qk,Fukunaga:2003sz,Cheng:2001ys,Redi:2016esr} or by choosing the axion decay constant sufficiently low. The latter option is already excluded in its most minimal form \cite{Raffelt:1985nk,Raffelt:1987yu,Lai:2002kf,Artamonov:2005ru,Abouzaid:2008xm,Ceccucci:2014oza}, though the constraints can be evaded in models which raise the axion mass without affecting the alignment of its potential with the CP preserving vacuum \cite{Rubakov:1997vp,Berezhiani:2000gh,Hook:2014cda,Fukuda:2015ana,Dimopoulos:2016lvn,Holdom:1982ex,Choi:1988sy,Holdom:1985vx,Dine:1986bg,Flynn:1987rs,Choi:1998ep,Agrawal:2017ksf,Agrawal:2017evu,Hook:2019qoh}. The collider phenomenology of such models maps directly onto the simplified model in \eqref{eq:ALPL} if the axion is of the KSVZ class \cite{Kim:1979if,Shifman:1979if}. In the case of Dine-Fischler-Srednicki-Zhitnitsky axions~\cite{Dine:1981rt,Zhitnitsky:1980tq} coupling to SM leptons are also present, even though the best sensitivity at these masses might still come from the diphoton final state~\cite{CMS:2012fgd,CMS:2019buh,LHCb:2018cjc,Cacciapaglia:2017iws,BuarqueFranzosi:2021kky}. 

Finally, the ALP in \eqref{eq:ALPL} could couple to the dark matter particle and be responsible for its freeze-out into SM states. A natural realization of this model with $\mathcal{O}(1)$ coupling constants predicts $f_a\sim$TeV \cite{CidVidal:2018eel}.

In the above, we have always assumed that the diphoton resonance is directly produced in the partonic collision, typically associated with a gluon. This need not to be the case however, and such a new state could also be produced in the decay of a heavier state, such as the Higgs. Specifically, if we consider a scalar field $S$ with couplings
\begin{equation}\label{eq:higgsimmodel1}
\mathcal{L}\supset \frac{\lambda_S}{2} S^2 H^\dagger H +\frac{\alpha_{\rm em}}{4\pi}\frac{1}{\Lambda_{\gamma\gamma}} S F^{\mu\nu}F_{\mu\nu}+\frac{\alpha_s}{4\pi}\frac{1}{\Lambda_{gg}} S G^{\mu\nu}G_{\mu\nu}\ ,
\end{equation}
where we can neglect possible couplings to the $WW$, $ZZ$ and $Z\gamma$ operators as long as $m_S\ll m_W$. This simplified model reproduces exotic Higgs decays of the form \mbox{$h\to 4\gamma$},  $h\to 2\gamma 2 g$ and  $h\to 4g$, whose relative importance depends on $\Lambda_{\gamma\gamma}/\Lambda_{gg}$. The four gluon channel is likely hopeless, but ATLAS and CMS have already performed searches for the $h\to 2\gamma 2 g$ \cite{ATLAS:2018jnf} and $h\to 4\gamma$ \cite{ATLAS:2015rsn,CMS-PAS-HIG-21-003} channels. These searches place the following constraints for $m_S\approx 20$ GeV
\begin{align}
&\mathrm{Br}(h\to SS)\times \mathrm{Br}(S\to \gamma\gamma)\times \mathrm{Br}(S\to gg)\lesssim 0.1\nonumber\\
&\mathrm{Br}(h\to SS)\times \mathrm{Br}(S\to \gamma\gamma)^2\lesssim 2\times 10^{-4}.
\end{align}
In the context of the simplified model in \eqref{eq:higgsimmodel1}, the $h\to4\gamma$ channel is therefore always more powerful and no trigger-level analysis in the diphoton topology is needed. 

The reason is the \mbox{$\mathrm{Br}(S\to gg)+\mathrm{Br}(S\to \gamma\gamma)\leq1$} constraint, which is however easily evaded by adding an additional scalar 
\begin{align}\label{eq:higgsimmodel2}\nonumber
\mathcal{L}\supset &\lambda_S S_1 S_2 H^\dagger H +\frac{\alpha}{4\pi}\frac{1}{\Lambda_{\gamma\gamma}} S_1 F^{\mu\nu}F_{\mu\nu}\\&+\frac{\alpha_s}{4\pi}\frac{1}{\Lambda_{gg}} S_2 \mathrm{Tr}\left[G^{\mu\nu}G_{\mu\nu}\right].
\end{align}
This model only induces the $h\to S_1S_2 \to gg \gamma\gamma$ topology. While we are not aware of a direct, top-down theory motivation for this specific topology, \eqref{eq:higgsimmodel2} is a perfectly plausible and fairly economical  extension of the SM Higgs sector. As such, it is important that it be covered as well as possible, since Nature may very well be more clever (or devious) than modern day model builders. The ATLAS search \cite{ATLAS:2018jnf} in particular makes use of the diphoton trigger path discussed in the introduction, which limits their reach to $m_{\gamma\gamma}\geq 20$ GeV. If a trigger-level analysis for lower invariant masses proves to be feasible, the $h\to 2\gamma 2 g$ topology should therefore benefit as well.

In addition to the above simple, two step cascade models more complicated dynamics can be realized in hidden valley models \cite{Strassler:2006im,Han:2007ae}, which can yield one or more low mass diphoton resonances \cite{Ellis:2012zp,Knapen:2021eip}.

\section{Discussion\label{sec:conclusions}}

Low mass diphoton resonances could hide in plain sight, being rejected by the experimental trigger selections. We showed that the uncharted mass window below the current reach of diphoton searches and above the existing constraints from flavor physics can be probed by ATLAS and CMS. While probing masses down to roughly 10 GeV is probably feasible with the current triggers, we find that a trigger-level analysis with modified isolation requirements may cover the whole mass region down to masses of order 1 GeV. 

We considered a simple, modified isolation requirement which was already used by ATLAS in a prior analysis~\cite{ATLAS:2015rsn} and argue that it could significantly increase the experimental acceptance for light resonances if implemented in a trigger-level analysis. Though the increased signal acceptance comes at the cost of a larger background rate, we nevertheless find otherwise inaccessible parameter space. That said, there are likely major technical challenges associated with a high rate, trigger-level analysis which can only be assessed to a satisfactory level by the collaborations themselves. The size of the fake photon background in particular should be verified in data before a definitive claim can be made.  

We remark that our analysis strategy is deliberately simplistic, and perhaps even naive, given the expected computational constraints on a trigger-level analysis. A number of more sophisticated analysis strategies have already been proposed for scenarios where the trigger is not the main challenge, most notably the topology $h\to 2a\to 4\gamma$ \cite{Dobrescu:2000jt,Draper:2012xt} but also cascade decays from heavier, exotic resonances \cite{Toro:2012sv}. Jet substructure and machine learning techniques applied to the energy depositions in the photon isolation cone appear to be promising \cite{Ellis:2012zp,Allanach:2017qbs,Sheff:2020jyw,Wang:2021uyb,Ren:2021prq}. Should it be possible to implement some of these techniques in a trigger-level analysis, they should yield substantially better sensitivity than our comparatively simplistic isolation requirement.  

A more ambitious possibility would be to push to even lower $m_{\gamma\gamma}$, to the extent that both photons merge and one must rely on the single photon L1 seed. This effectively involves modifying the photon identification requirements themselves. This was already accomplished in an (offline) ATLAS  \cite{ATLAS:2018dfo} search for a pair of photon jets.
However the pay-off could be very substantial if this can also be done for a single photon jet as a trigger-level analysis, given that the single photon trigger jumps from \mbox{22 GeV} to 120 GeV between L1 and HLT \cite{ATLAS:2019dpa}. While the technical challenges should not be underestimated, we conclude that trigger-level searches for low mass diphoton resonances and photon jets could open qualitatively new parameter space for the discovery of low mass diphoton resonances.



\begin{acknowledgements}
 We thank Liron Barak, Matt Low, Alberto Mariotti, Marco Montella,  Jos\'e Ocariz, Michele Papucci, Luis Pascual, Dean Robinson, Filippo Sala and Kohsaku Tobioka for many discussions, comments on the draft and earlier work on similar topics. We thank Christian Bauer, Zoltan Ligeti, Michelangelo Mangano and Nicholas Rodd in particular for discussions on the details of the photon fragmentation functions.  We are especially grateful to Caterina Doglioni for the discussions that initiated this project and for detailed comments on the manuscript. SKn was supported by the Office of High Energy Physics of the U.S. Department of Energy under contract DE-AC02-05CH11231. SKu was supported in part by the National Science Foundation (NSF) grant PHY-1915314 and the U.S. Department of Energy Contract DE-AC02-05CH11231.  SKu thanks Aspen Center of Physics, supported by NSF grant PHY-1607611, for its hospitality while this work was in progress. All authors thank the Galileo Galilei Institute for Theoretical Physics for its hospitality  and the INFN for partial support during the initial and final phases of this work. 
\end{acknowledgements}

\appendix

\section{Simulation details}\label{app:appendix}
\subsection{Background simulation\label{app:background}}
\begin{figure}
    \centering
    \includegraphics[width=\columnwidth]{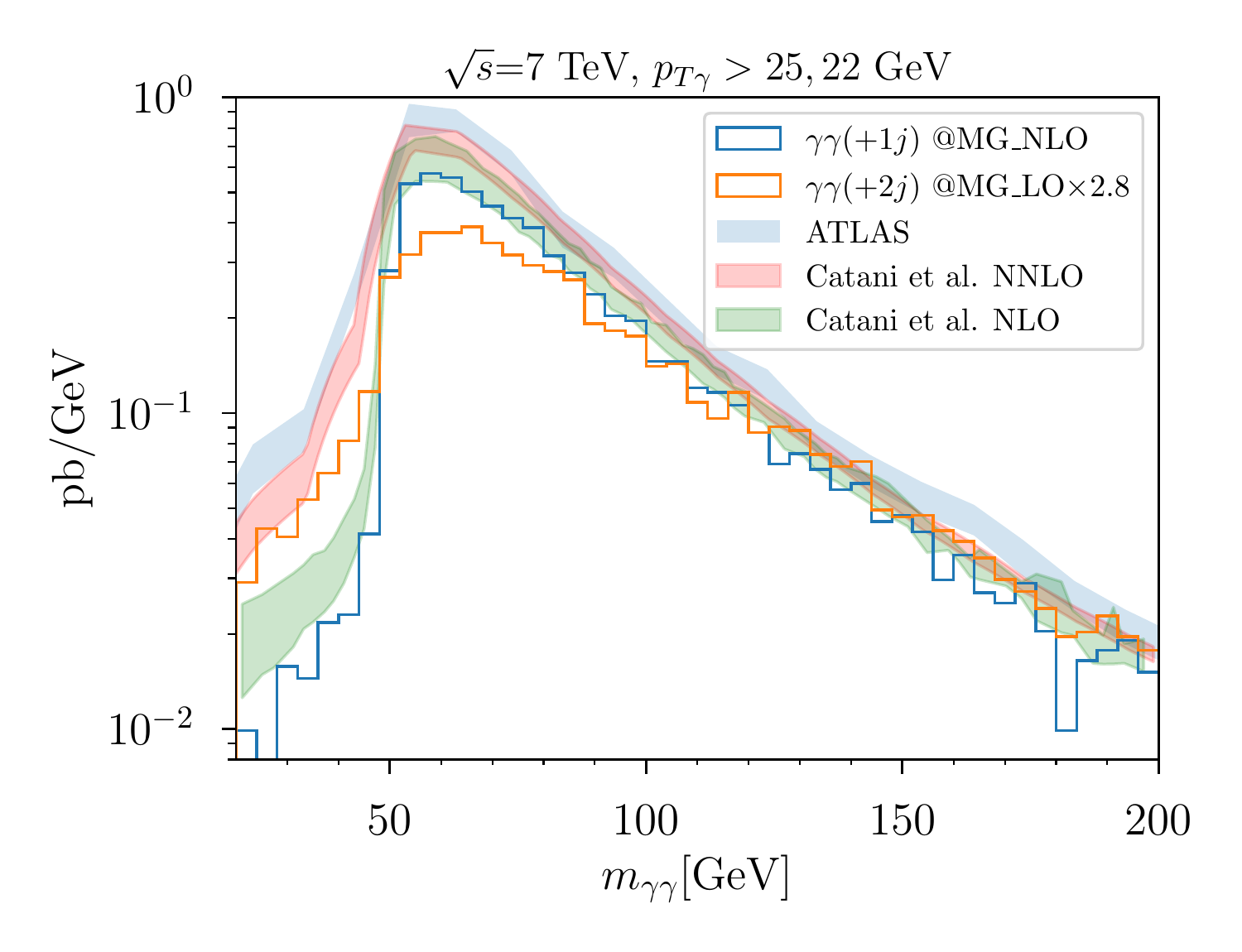}
    \caption{Diphoton invariant mass plot from Catani~et~al.~\cite{Catani:2018krb}, overlaid with our simplified procedure (orange, rescaled by an NLO K-factor of 2.8) and Madgraph@NLO (blue). See text for details. 
    }
    \label{fig:catani_compare}
\end{figure}

To validate our simulation framework, we compare with the NNLO calculations performed by Catani~et~al.~\cite{Catani:2011qz,Catani:2018krb}, which have shown good agreement with ATLAS data at 7 TeV \cite{ATLAS:2012fgo}. This comparison is shown in Fig.~\ref{fig:catani_compare}. Note that these results are obtained with a fixed isolation threshold $E_T^{\rm MAX}=2$~GeV, rather than with the sliding threshold we used in \eqref{eq:isoATLAS} and \eqref{eq:modiso}. The leading and subleading photon must satisfy $p_T>25$ GeV and $p_T>22$ GeV respectively, as well as a rapidity cut $|y_\gamma|<2.37$. At leading order, the process is just the $q\bar q\to \gamma\gamma$, for which there is no phase space for $m_{\gamma\gamma}<50$ GeV. At NLO, new processes contribute such as $q g\to \gamma\gamma q$ and $q g\to \gamma q$, where a second photon is supplied by the quark fragmentation function. This contribution is enhanced by the parton distribution functions (pdf's), despite being formally higher order than the $q\bar q\to \gamma\gamma$ process. The NLO calculation also contains virtual corrections to $q\bar q\to \gamma\gamma$.  The additional jet at NLO opens up the phase space for $m_{\gamma\gamma}<50$ GeV. Finally, the NNLO calculation allows for an additional soft jet and includes virtual corrections via the box diagrams. The latter are important because they include the $gg\to \gamma\gamma$ process, which is further pdf enhanced compared to the LO and NLO contributions. The blue curve, diphoton events matched up to one jet, in Fig.~\ref{fig:catani_compare} is obtained with Madgraph@NLO and is in good agreement with the Catani et~al.~NLO calculation.

\begin{figure*}
    \centering
    \includegraphics[width=0.33\textwidth]{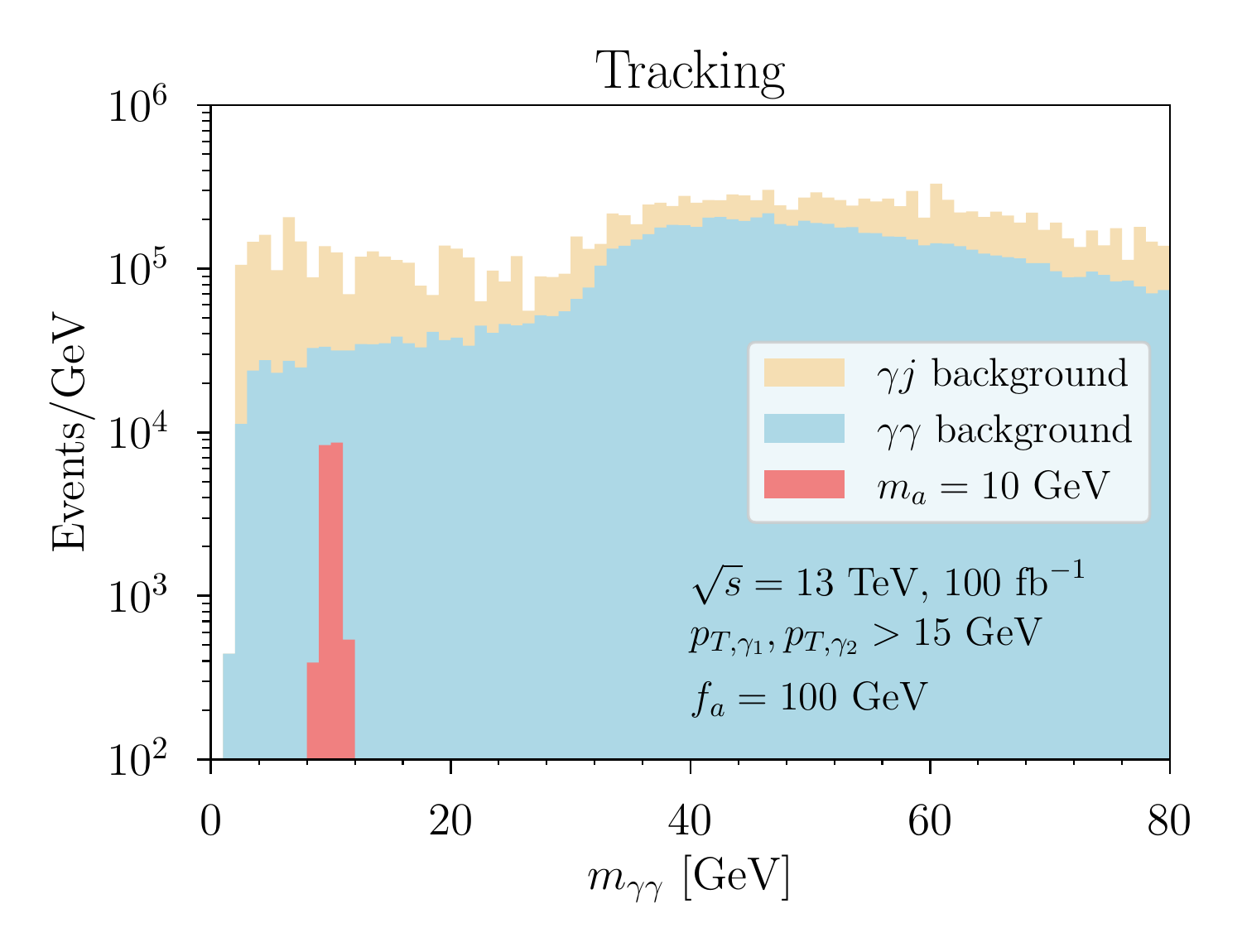}\hfill
    \includegraphics[width=0.33\textwidth]{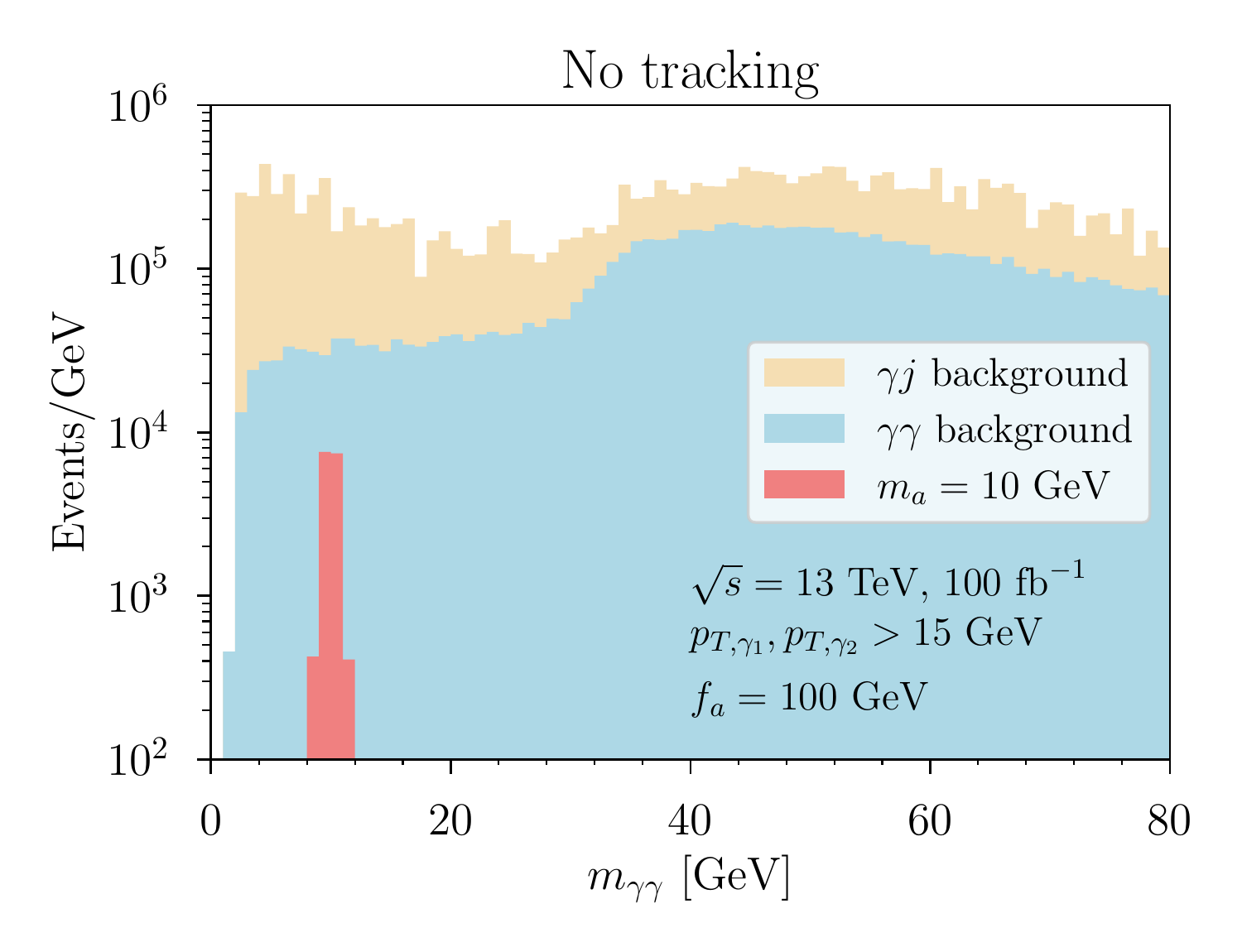}\hfill
    \includegraphics[width=0.33\textwidth]{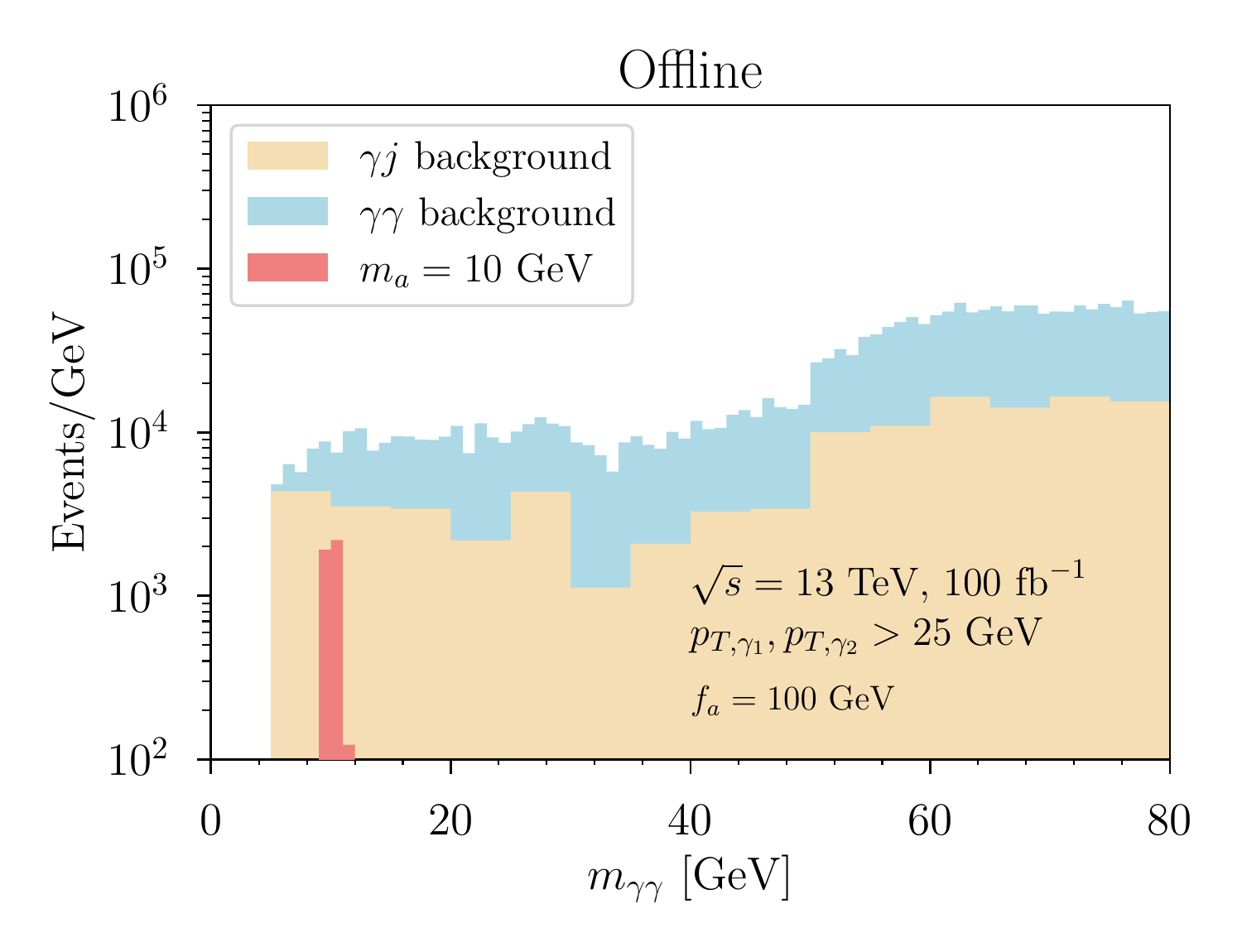}
    \caption{Stacked invariant mass distribution for the signal and the background for the three scenarios described in Sec.~\ref{sec:analysis}, for a larger mass range. A representative signal resonance is shown for $m_a=10$~GeV with total number of events corresponding to $f_a=100$~GeV.
 }
    \label{fig:inv_mass_dist_app}
\end{figure*}

For computational reasons, we used a leading order matched sample, rescaled to the inclusive NLO cross section,  as described in Sec.~\ref{sec:background}. To estimate the error this simplification introduces, we apply it to the kinematical selections used by Catani~et~al., as is shown by the orange curve in Fig.~\ref{fig:catani_compare}. Here we neglected photons from the non-perturbative part of the fragmentation, which is a subleading component for standard isolation (see the rightmost panel of Fig.~\ref{fig:inv_mass_dist}). Our simplified procedure somewhat underestimates the peak but is in $\mathcal{O}(1)$ agreement with the low $m_{\gamma\gamma}$ tail. While this may seem far from ideal, it worth noting that the sensitivity projection in Fig.~\ref{fig:projections} only scale as the $4^{\text{th}}$ root of the background. Our qualitative conclusions are therefore not very sensitive to this level of background mismodeling. For completeness we extend the invariant mass distributions in Fig.~\ref{fig:inv_mass_dist} to the full mass range in Fig.~\ref{fig:inv_mass_dist_app}. We clearly see the feature around twice the $p_T$ threshold, consistent with the discussion in the previous paragraph.


Finally, we also comment on the dependence of the hadronic contribution to the photon background on the choice of isolation criterion. For this purpose, we simulated a large $q\bar q$ sample with \texttt{Pythia 8} and quantify the probability of a jet producing an isolated photon from a hadronic decay.  To cleanly separate this contribution from the perturbative part of the fragmentation function we veto all electromagnetic emissions during the parton shower. (These contributions were discussed separately in Sec.~\ref{sec:background}, specifically in Fig.~\ref{fig:vetocone}.)

We find that with the modified isolation criterion, the fake rate increases by a bit more than an order of magnitude. Using tight isolation does reduce the background somewhat, but does not fully compensate for the aforementioned increase. That said, given the rather special corner of phase space that is being probed here, a data-driven approach will be essential to validate the Monte Carlo predictions.

\begin{figure}[t]
    \centering
    \includegraphics[width=\columnwidth]{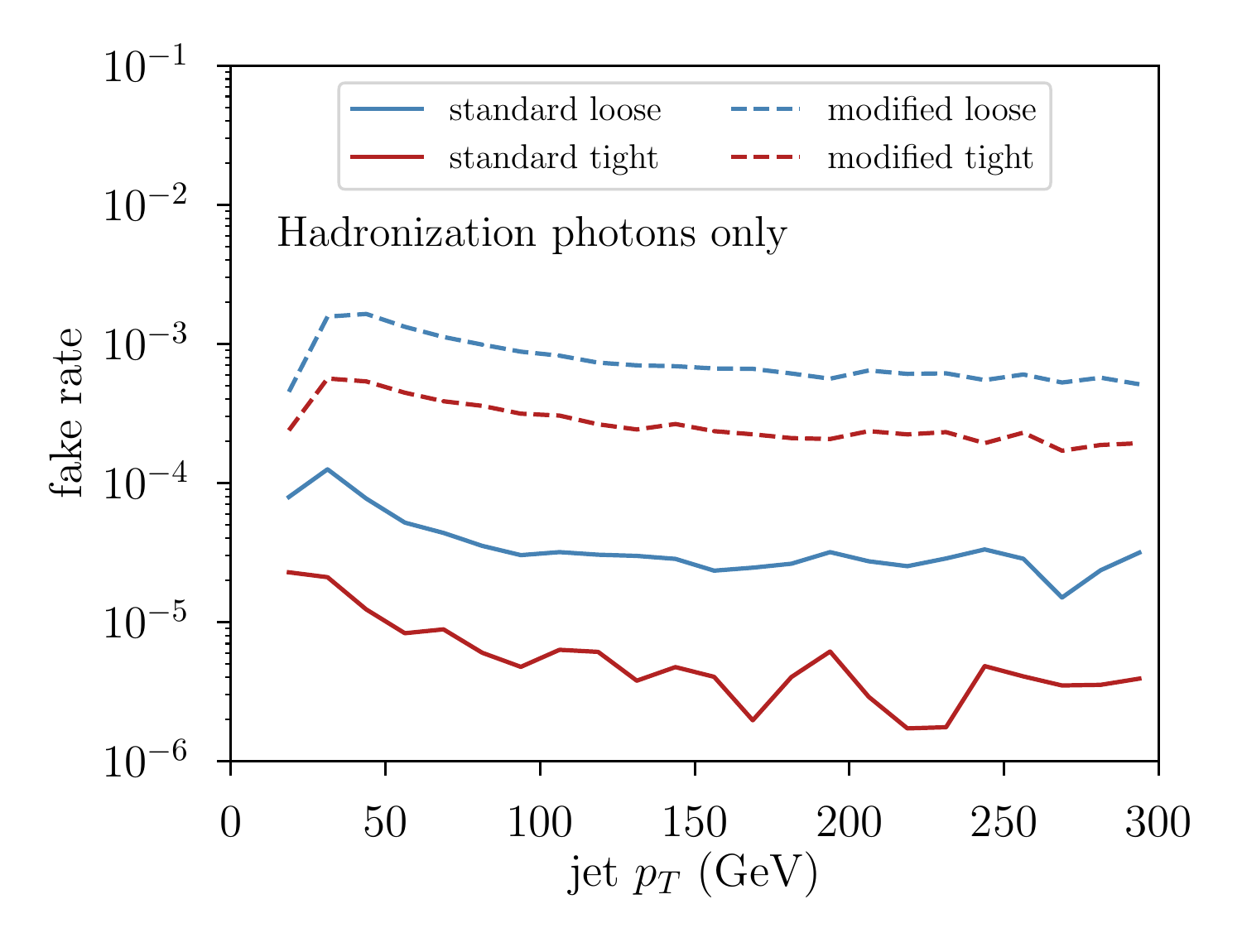}
    \caption{Probability of a quark jet fragmenting into an isolated photon with $p_T>15$ GeV, for standard and modified isolation, with loose and tight selections (see Sec.~\ref{sec:iso}). Only photons from hadronic decays are considered. For photons from the perturbative part of the parton shower we refer to Fig.~\ref{fig:vetocone}.} 
    \label{fig:hadronic_fake_rate}
\end{figure}


\subsection{Mass resolution\label{app:resolution}}
\begin{figure}
    \centering
    \includegraphics[width=\columnwidth]{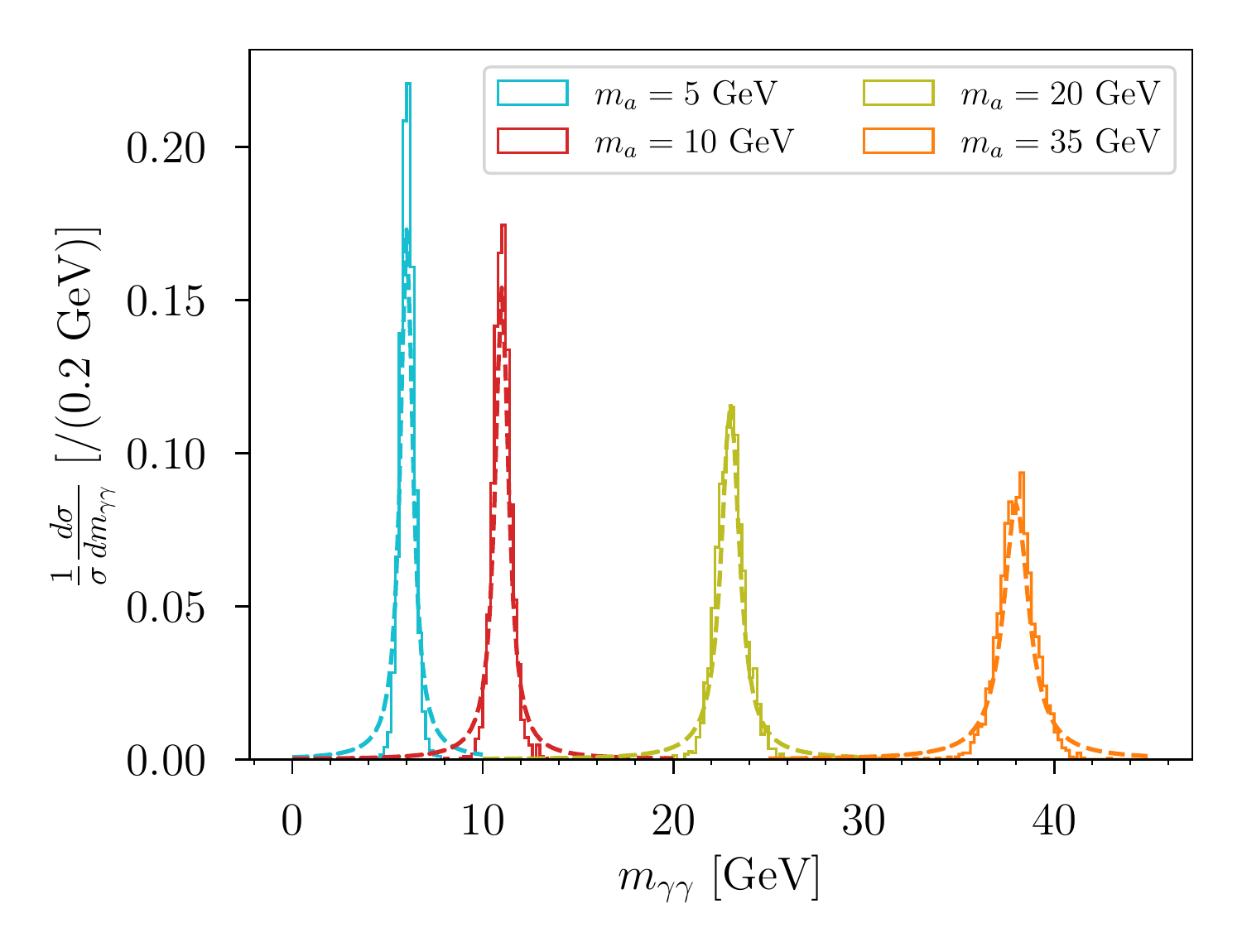}
    \caption{Mass resolution for a diphoton resonance, as reproduced by \texttt{Delphes 3} with the standard ATLAS configuration card (solid). To determine the width of the overlaid Lorentzian line shape we take a $p_T$ threshold of 15~GeV (dashed), as described in the text. } 
    \label{fig:mass_res}
\end{figure}

Given the limited number of handles at our disposal, the main background discrimination comes from hunting for a bump on top of a large, background. The resolution on the invariant mass is therefore very important. This is handled automatically by \texttt{Delphes 3}, as shown in Fig.~\ref{fig:mass_res}. Given its importance, we seek to verify the built in \texttt{Delphes 3} parametrization explicitly. Concretely, the energy resolution of the ATLAS ECAL is parametrized by \cite{ATLASElectromagneticBarrelCalorimeter:2006ymk}
\begin{equation}\label{eq:enres}
\frac{\delta E_\gamma}{E_\gamma}= 10\% \left(\frac{\text{GeV}}{E_\gamma}\right)^{1/2}\oplus 0.7\%.
\end{equation}
where the $\oplus$ notation indicates that the errors are added in quadrature. 
The mass resolution is given by
\begin{equation}\label{eq:massresolution}
\frac{\delta m_{\gamma\gamma}}{m_{\gamma\gamma}} = \frac{1}{2}\left(\frac{\delta E_{\gamma,1}}{E_{\gamma,1}}\oplus \frac{\delta E_{\gamma,2}}{E_{\gamma,2}}\right)\oplus \frac{0.0247}{\Delta R_{\gamma\gamma}}
\end{equation}
where for the angular term we have taken the off-line granularity of the calorimeter. This is slightly worse than the true off-line resolution of the ECAL but better than the L1 angular resolution. The assumption here is that the fine grained information for the ECAL tower can be written out for a small region of interest around the candidate resonance. 
To compare with Fig.~\ref{fig:mass_res}, we estimate $\delta m_{\gamma\gamma}$ by taking $\eta\approx 0$ for both photons and saturating $E_{\gamma,1}$ and $E_{\gamma,2}$ to the $p_T$ thresholds. This determines the energy resolution via~\eqref{eq:enres}. We then solve for $\Delta R_{\gamma\gamma}$ in~\eqref{eq:massestimate} by using the injected invariant mass. The resulting numbers are plugged into \eqref{eq:massresolution} to obtain the estimated mass resolution. We overlaid a Lorentzian line shape with width $\delta m_{\gamma\gamma}$ to visualize the agreement with the full \texttt{Delphes 3} simulation.

%
%
%
%

\bibliographystyle{JHEP}
\bibliography{alp_photon}

\end{document}